\documentstyle[psfig]{mn}
\newif\ifAMStwofonts

\def\eso36{\hbox{ESO~3.6-m}}
\def\ni{\noindent}

\title[A 1.4 GHz Survey in the Southern ELAIS Region] {A 1.4 GHz Survey of
  the Southern ELAIS Region} 

\author[C. Gruppioni, P. Ciliegi, M.  Rowan-Robinson et al.]
{\parbox[]{6.5in} {C.  Gruppioni$^{1}$\thanks{e-mail:
      c.gruppioni@ic.ac.uk}, P.  Ciliegi$^{2,3}$, M. Rowan-Robinson$^{1}$,
    L. Cram$^{4}$, A.  Hopkins$^{4}$,
C.~Cesarsky$^5$, 
L.~Danese$^6$,  
A.~Franceschini$^7$,
R.~Genzel$^8$,  
A.~Lawrence$^9$, 
D.~Lemke$^{10}$,  
R.G.~McMahon$^2$, 
G.~Miley$^{11}$,
S.~Oliver$^1$,  
J-L.~Puget$^{12}$,  
B.~Rocca-Volmerange$^{13,12}$ } \\ 
$^1$ Astrophysics Group, Imperial College London, Prince Consort Road, London 
SW7 2BZ, U.K.\\
$^2$ Institute of Astronomy, Madingley Road, Cambridge, CB3 0HA \\
$^3$ Osservatorio Astronomico di Bologna, via Zamboni 33, Bologna, I-40126,
Italy \\
$^4$ School of Physics, University of Sydney, NSW,
2006, Australia \\
$^5$ Service d'Astrophysique, Saclay, 91191, Gif-sur-Yvette, Cedex,
France \\ 
$^6$ SISSA, Via Beirut 2--4, Trieste, Italy \\
$^7$ Osservatorio Astronomico di Padova, Vicolo dell'Osservatorio 5,
I--35122, Padova, Italy \\
$^8$ Max-Planck-Institut f\"ur Extraterrestrische Physik,
Giessenbachstrasse, D-8046, Garching bei M\"unchen, Germany \\
$^9$ Institute for Astronomy, University of Edinburgh, Blackford Hill,
Edinburgh, EH9 3HJ \\
$^{10}$ Max Plank Institute, Heidelberg, Germany \\
$^{11}$ Leiden Observatory, P.O. Box 9513, 2300 RA, Leiden, The Netherlands \\
$^{12}$ Institut d'Astrophysique Spatiale, Bat. 121, Universit\'e Paris XI,
F-91405 Orsay Cedex, France \\
$^{13}$ Institut d'Astrophysique de Paris, CNRS, 98 bis Bd. Arago,
F-75014, Paris, France} 

\date{Accepted ?? Received ??}

\pagerange{\pageref{firstpage}--\pageref{lastpage}} 

\pubyear{1997}

\def\LaTeX{L\kern-.36em\raise.3ex\hbox{a}\kern-.15em
  T\kern-.1667em\lower.7ex\hbox{E}\kern-.125emX}

\begin{document}
 
\label{firstpage}

\maketitle
 
\begin{abstract}
  A deep survey of the European Large Area ISO Survey (ELAIS) field in the
  southern celestial hemisphere (hereinafter S1) has been carried out with
  the Australia Telescope Compact Array (ATCA) at 1.4 GHz. The S1 region,
  covering about 4 square degrees, has been also surveyed in the mid- and
  far-infrared (5-200 $\mu$m) with the Infrared Space Observatory (ISO).
  The radio survey provides uniform coverage of the entire S1 region, with
  a sensitivity ($5\sigma$) of 0.4 mJy over the whole area and 0.2 mJy in
  the center. To this sensitivity, virtually all the radio counterparts of 
  the far-infrared
  extragalactic ISO sources should be detected. This paper presents a radio
  sample--complete at the 5$\sigma$ level--consisting of 581 sources
  detected at 1.4 GHz. Of these radio sources, 349 have peak flux density
  in the range 0.2-1 mJy, forming a new homogeneous sample of sub-mJy
  radio sources. Due to its size, depth and multi-waveband coverage, the
  sample will allow us to study in greater detail the sub-mJy radio source
  population.

  The full catalogue will be available from http://athena.ph.ac.uk/ 
\end{abstract}


\section{Introduction}
The European Large Area ISO Survey (ELAIS, Oliver et al. 1997; Oliver et
al.  1998 in preparation) is a collaboration between 20 European
institutes.  It involves a deep, wide-angle survey at high galactic
latitudes, at wavelengths of 6.7 $\mu$m, 15 $\mu$m, 90 $\mu$m and 175 $\mu$m
with the Infrared Space Observatory (ISO).  The 6.7 $\mu$m and 15 $\mu$m
surveys were carried out with the ISO-CAM camera (Cesarsky et al. 1997) to
$5\sigma$ sensitivities of $\sim$0.6 and 2 mJy, respectively. The 90 $\mu$m
and 175 $\mu$m surveys were carried out with the ISO-PHOT photometer (Lemke
et al. 1994), to a $5\sigma$ sensitivity of $\sim$60--80 mJy at 90 $\mu$m.
With a sensitivity of $\sim$60 mJy at 90 $\mu$m, ELAIS will be the deepest
far-infrared survey performed with ISO. The survey has detected objects
5--10 times fainter than IRAS in the 50--100 $\mu$m range, and 20--50 times
fainter than IRAS in the 10--20 $\mu$m range.  The survey is divided into 4
fields (one of which, S1, is in the southern hemisphere), and covers a
total area of $\sim$ 13 deg$^{2}$ at 15 $\mu$m and 90 $\mu$m, $\sim$7
deg$^{2}$ at 6.7 $\mu$m and $\sim$3 deg$^{2}$ at 175 $\mu$m.

ELAIS will allow us to study dust emission in normal galaxies and the
evolution of star formation to high redshifts, testing competing scenarios
for the formation of elliptical galaxies, exploring the IRAS galaxy
population to higher $z$, and possibly unveiling new classes of objects or
unexpected phenomena.  Since dust plays an important role in most of these
goals, and since many of the thousands of galaxies detected in the ELAIS
survey will be at high redshift and probably obscured in the optical bands,
radio observations will play a crucial role in assessing the reliability of
the ELAIS source list and in facilitating source identification in faint or
empty optical fields. Additionally, radio data will be important in the
optical identification phase because the spatial resolution of ISO is
insufficient to identify unambiguously many of the faintest ISO sources.
Even at 15 $\mu$m the survey resolution is $\sim$10 arcsec, while at 90
$\mu$m it is almost 1 arcmin, so there are multiple optical candidates
within each error box.

The three northern ELAIS fields (N1, N2 and N3) have been surveyed in the
radio band (at 1.4 GHz) with the Very Large Array (VLA) down to an average
flux density limit of 0.25 mJy (5$\sigma$) over a total area of 4.22
deg$^2$.  Details of the VLA observations and a description of the
catalogue containing 867 radio sources can be found in Ciliegi et al.
(1998).

In this paper we describe radio observations of the ELAIS field located in
the southern celestial hemisphere, S1. This field, centered at
$\alpha$(2000) = 00$^h$ 34$^m$ 44.4$^s$, $\delta$(2000) = -43$^{\circ}$
28$^{\prime}$ 12$^{\prime \prime}$, covers an area of the sky of about
$2^{\circ} \times 2^{\circ}$.  We observed the whole area at 1.4 GHz with
the Australia Telescope Compact Array (ATCA) in the 6-km configuration
(maximum baseline length), with a resolution of $8 \times 15$ arcsec. We
obtained uniform radio coverage over the whole S1 region, with sufficient
sensitivity ($1\sigma \simeq 80~ \mu$Jy) to detect virtually all radio
counterparts of the ISO galaxies.

By combining these radio observations with the available ISO data, we will
investigate the radio/far-infrared correlation (Helou, Soifer \&
Rowan-Robinson 1985) in star-forming galaxies to a flux density
significantly deeper than that reached by IRAS.  When spectral informations
become available for the optical counterparts of the ISO and ATCA sources
in the field, we will investigate the trivariate IR-radio-optical
luminosity function and its evolution. This study will be of fundamental
importance for interpreting the source counts at different wave-bands and
for elucidating contributions from different classes of objects. Moreover,
we will be able to determine the influence of obscuration on the inference
of the star-formation rate, and to constrain the star formation history of
the Universe to $z \simeq$ 1.0--1.5 with two complementary samples, selected
in the infrared and radio, within the same volume of the Universe (see Cram
et al. 1998; Oliver, Gruppioni \& Serjeant 1998).

The radio sample is also important in its own right, since it constitutes
one of the largest homogeneous samples of sub-mJy radio sources, whose
nature is still uncertain.  Previous work has suggested that the sub-mJy
radio source population is composed largely of star-burst galaxies at
moderate redshifts (i.e. Windhorst et al. 1985; Benn et al. 1993).
However, there is little information about the nature and the true redshift
distribution of the sub-mJy population, because only a small fraction of
them have been optically identified and have measured redshifts (Benn et
al. 1993; Rowan-Robinson et al. 1993).  Although it is commonly believed
that most of the sub-mJy radio sources are associated with star-forming
galaxies, recent extensions to fainter optical magnitudes (and higher
identification percentages) hint strongly at an increasing fraction of
early-type galaxies (Gruppioni, Mignoli \& Zamorani 1998). This suggestion,
together with apparently discordant results from optical identifications of
even fainter $\mu$Jy radio samples ( Windhorst et al.  1995; Hammer et al.
1995; Richards et al. 1998) complicate the picture.  The ATCA survey in the
ELAIS field S1, together with the VLA surveys of the northern fields N1, N2
and N3 (Ciliegi et al.  1998) will shed new light on the sub-mJy population
and its connection to the faint blue and infrared star-burst galaxies. The
sensitivity, areal coverage and multi-wavelength character (optical, near-,
mid- and far-infrared and radio) of the ELAIS project will ensure that
it makes an important contribution to all of these problems.

In section 2 of the paper we describe the radio observations. Section 3
discusses the data reduction strategy and section 4 presents the source
catalogue. In section 5 we discuss the radio source counts and present our
conclusions in section 6.

\section{Radio Observations}

\subsection{Observing Strategy}
By choosing an observing frequency of 1.4 GHz, we ensure a large
instantaneous field of view ($33$ arcmin primary beam) and hence shorten
the observing time necessary to cover the entire S1 area. The sensitivity
is also higher for the same integration time with respect to higher
available frequencies (2.4, 4.86 and 8.44 GHz). Moreover, a frequency of
1.4 GHz reveals well any recent star-formation activity, seen at this
frequency as synchrotron radiation excited by supernova remnants (Condon
1992).

We observed in continuum mode with a bandwidth $B$ of 128-MHz.  The
sensitivity $\Delta S$ (mJy) is

\begin{equation}
  \Delta S~ = 0.0755 S_{sys} (NtB)^{-1/2} \label{eq:sens}
\end{equation}

\ni where $S_{sys}$ (in Jy) is the system sensibility (350 Jy at 1.4 GHz),
$t$ is the integration time (in minutes), $B$ is the bandwidth (in MHz) and
$N$ is the number of baselines.

\subsection{The Mosaic Technique}

The ATCA is an east-west array which uses the earth's rotation to sample
visibilities over elliptical loci in the spatial frequency ($u,v$) plane.
Its control system allows imaging over a field or a source whose extent is
larger than the primary beam by cycling through a grid of pointings on the
sky, recording the visibilities for each pointing intermittently. This {\it
  mosaic observing mode} is an efficient way to obtain uniform quality
imaging over a large field, and has been used to construct our survey.

The grid of pointings in the mosaic was designed to yield a homogeneous
radio source sample. We aim to have a detection threshold independent of
source position, and therefore require uniform noise over the area of
interest. With the mosaic technique, images obtained with single pointings
are combined together into a large image (mosaic) of the entire observed
region. We used `linear combination' mosaicing rather than the `joint
deconvolution' approach (described by Sault, Staveley--Smith and Brouw
1996) because the latter requires greater computing resources and is less
suited to images dominated by compact sources.  Linear combination
mosaicing takes a pixel-by-pixel weighted mean of the single pointings,
where the weights are determined from the primary beam response and the
noise level in that pixel (Sault \& Killeen 1995).  The value of each pixel
in the final mosaic depends on the pointing configuration and, in
particular, on the grid spacing.  For our survey, an optimal trade-off
between uniformity of sensitivity and efficiency of telescope usage was
obtained by adopting a rectangular grid of $7 \times 7$ fields, with a
regular spacing between fields equal to $\theta_{FWHP} /\sqrt{2} \sim 20$
arcmin (where $\theta_{FWHP}$ is the full width at half power of the
primary beam, $33$ arcmin at 1.4 GHz). Figure 1 shows the sky position and
orientation of the S1 region. Circles drawn with a radius of $20$ arcmin
show the 49 ATCA pointings.

\begin{figure}
\centerline{\psfig{figure=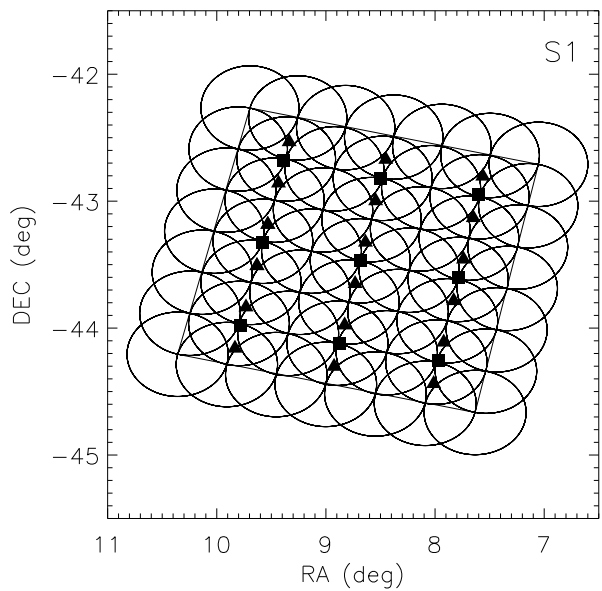,width=9cm}} 
 \footnotesize
\caption{
  The sky position and orientation of the ISO S1 survey region. The outer
  rectangle is $2^{\circ} \times 2^{\circ}$.  The filled triangles show the
  ISO-CAM 15 $\mu$m pointings, while the filled squares show the ISO-PHOT
  90 $\mu$m pointings.  The circles (drawn with a radius of $20$ arcmin)
  show the ATCA mosaic pointings.}
\label{fig_elais_grids_status}
 
\end{figure}
 
\subsection{Observations}
The ATCA observations of the ELAIS field S1 were performed on the nights of
1997 June 23, 24, 26, 27 and 28, in runs of 12 hours per night. The
theoretical sensitivity is 80 $\mu$Jy ($1\sigma$) over the whole area.
Since the far-infrared and radio luminosities of normal galaxies are
tightly correlated (Helou, Soifer \& Rowan--Robinson 1985; Condon 1992),
flux-limited samples of normal galaxies selected at far-infrared and radio
wavelengths at the corresponding sensitivity are nearly identical. Given a
$5\sigma$ sensitivity of 60 mJy for ISO observations at 90 $\mu$m, the
far-infrared/radio correlation predicts that 1.4 GHz radio counterparts of
the faintest ELAIS sources should be detected at a flux density a little
below 0.5 mJy.  Thus, the flux density limit reached by our observations,
corresponding to $5\sigma \simeq 0.4$ mJy, should allow us to detect
essentially all the ELAIS ISO-PHOT and most of the ISO-CAM galaxies.

To optimize ($u,v$) coverage we organised the observations by dividing
the 49 fields in four blocks of 13 fields each and observing each block for
12h + 3h (one night plus one quarter of the last night).  During this time
the 13 fields were observed with dwell times of 20 seconds each, adding 3
minutes for observing the secondary calibrator every hour (i.e. every 15
cycles of 13 fields).  The field corresponding to the center of the S1 area
was visited in each block, to obtain deeper sensitivity in the central
$\sim 20$ arcmin of the ELAIS field.

The primary flux density calibrator was PKS B1934-638, whose flux densities at
different frequencies are incorporated directly in the calibration
software.  The primary calibrator was observed for 10 min at the beginning
of each 12h observing run. The source PKS B0022-423 was used as a phase and
secondary amplitude calibrator.

Before observing we checked for the presence of strong radio sources in the
field, which could eventually compromise the sensitivity of the survey.
Although the ELAIS S1 field was not selected with radio observations in
mind, fortunately there are only a few bright radio sources within it, so
that the target noise level was obtained without serious problems of
confusion.  The Parkes-MIT-NRAO (PMN) Southern Survey (Griffith \& Wright
1993) reveals one bright radio source (PMN J0042-4413) just outside the
field, with a flux density of 1.2 Jy at 5 GHz.  There are also one 0.4 Jy
and one 0.2 Jy PMN source in the field, and several with lower flux
densities. We have observed the whole S1 area, expecting that not more than
a small part of the mosaic might be degraded by the sidelobes of strong
sources in or just outside the field.

We used both receivers at 1.4 GHz, each with a bandpass of 128 MHz. Using
the ATCA in this mode makes it hard to avoid strong interference in some
channels for at least part of the observation, but the doubled bandwidth
improves the sensitivity by a factor of $\sqrt{2}$.  The optimal
positioning of the bands was obtained by centering one frequency at 1.344
GHz (1.280-1.408 GHz) and the other at 1.472 GHz (1.408-1.536 GHz).

\section{Data Reduction}
The data were analyzed with the software package {\sc miriad}
(Multi-channel Image Reconstruction, Image Analysis and Display), which is
standard software for radio-interferometric data reduction specifically
adapted to the ATCA.

\begin{figure*}
%
%
%
\vspace{15cm}

\caption[S1 ATCA mosaic]{Grey scale image of the total 1.4 GHz mosaic in the ELAIS S1 field. The square region corresponds to the area observed by ISO.}
\label{fig_mosaic}

\end{figure*} 
 
In particular, {\sc miriad} provides {\it multi-frequency synthesis} ({\it
  mfs}) algorithms (Sault \& Wieringa 1994), which give the opportunity of
producing images with improved ($u,v$) coverage by combining accurately
the visibilities of individual channels.  Multi-frequency synthesis may be
used with the ATCA because, in continuum mode, the correlator provides a
bandwidth of 128 MHz subdivided into 32 frequency channels (of 4 MHz each).
The division of the wide passband into sub-channels reduces the effects of
bandwidth smearing.  Moreover, {\it mfs} fills in the ($u,v$) plane
despite the modest number of ATCA antennas, since different observing
frequencies produce different spatial frequency intervals $\Delta u$ and
$\Delta v$.

Before calibrating the data, we flagged all baselines and fringe
visibilities affected by any problems encountered during the observation
(such as antenna slewing intervals, interference, etc) using the tasks {\sc
  blflag}, {\sc tvflag} and {\sc uvflag}. Primary calibrator data were
flagged before calibrating, and the secondary calibrator after application
of the band-pass and instrumental polarization calibrations obtained from
the primary calibrator. Finally, the fully calibrated single field data
were flagged.  Each bandpass (1.344 and 1.472 GHz) was calibrated
separately, according to the standard calibration technique (Sault and
Killeen 1995).

The complete imaging procedure (described in detail by Prandoni 1997)
comprises imaging itself (performed with the {\sc miriad} task {\sc
  invert}), and several cycles of cleaning (task {\sc clean}) and
self-calibration (task {\sc selfcal}).  The whole procedure was applied
separately to each frequency to account for spectral variations of each
source, since flux density differences due to the intrinsic spectral slope
could be significant over the frequency interval of $2 \times 128$ MHz.
For each field we produced individual images with the same geometry and the
same celestial coordinate as reference point, to be combined together into
a single mosaic at the end of the reduction phase.

The images are $1200 \times 1200$ pixels square, and for each field a $2400
\times 2400$ pixels square beam was produced. An average synthesized beam
of $15 \times 8$ arcsec$^2$ was created for all the images.  Cell sizes of
$2.5$ arcsec per pixel were used, to obtain good re-sampling of the ($u,v$)
plane (at least 3 pixels within each dimension of the synthesized beam).
Natural weighting was used, to increase the sensitivity at the price of a
decrease in spatial resolution.  The images for each pointing center were
then {\sc clean}ed.  A few preliminary deconvolution iterations (performed
to find the model components for the self-calibration) were followed by a
self-calibration phase and then a deeper cleaning phase.  This cycle was
repeated twice. After subtracting from the visibility file the components
found by the clean algorithm, we flagged any residual visibilities still
affected by interference. Then we performed the deconvolution on the new
visibility data-set (after having recombined the previously subtracted
components).  Self-calibration was used to make additional corrections to
the antenna gains as a function of time, to improve the image quality and
to account for the fact that interpolated calibrator gains do not determine
the antenna gains perfectly at each time step.

To minimise the side-lobes of bright off-field sources, we performed a few
iterations of the cleaning algorithm on a double-sized (four times in area)
low resolution image of the field, subtracting the components found in the
external parts of the image from the visibility file.  The final images were
obtained by adding the components found by the deconvolution algorithm to
the residual image. This was done with the task {\sc restor}.  Once good
quality images of each field were obtained, we constructed the final
mosaics using the linear combination task {\sc linmos} (which also corrects
for primary beam attenuation).  Two mosaics were first created for each of
the two frequency bands and then added, resulting in a square $2^{\circ}
\times 2^{\circ}$ mosaic with almost uniform sensitivity over the full
region.  Figure~\ref{fig_mosaic} exhibits a grey scale image of the total
mosaic of the entire S1 area.

\begin{figure} 
\centerline{\psfig{figure=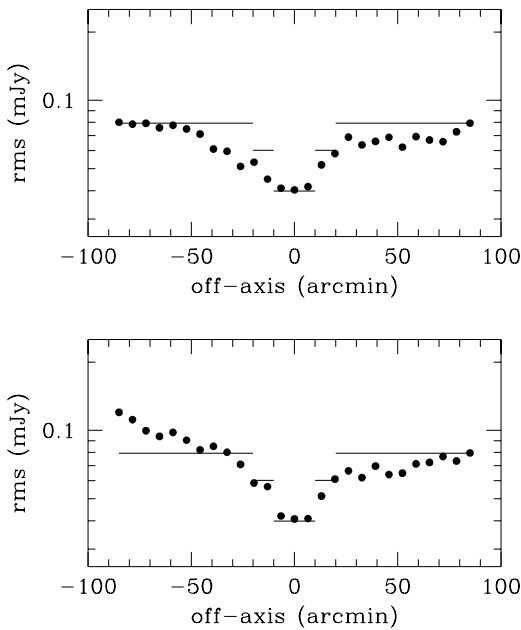,width=9cm} }
 
  \footnotesize
\caption[rms noise]{The rms noise as a function of distance from the field
  center, measured along the two diagonals of the square defined by the S1
  field.  {\it Top}: rms noise along the diagonal degraded by the sidelobes
  of PMNJ0042-4413; {\it Bottom:} rms noise along the diagonal not affected
  by the presence of the strong radio source. The solid lines
  (corresponding to 1$\sigma$ rms noise values of 80, 60 and 40 $\mu$Jy for
  $r > 20$ arcmin, $10 < r < 20$ arcmin, and $r < 10$ arcmin, respectively)
  represent the limiting fluxes assumed for the source extraction.}
\label{fig_noise}
\end{figure} 
\begin{figure} 
\centerline{\psfig{figure=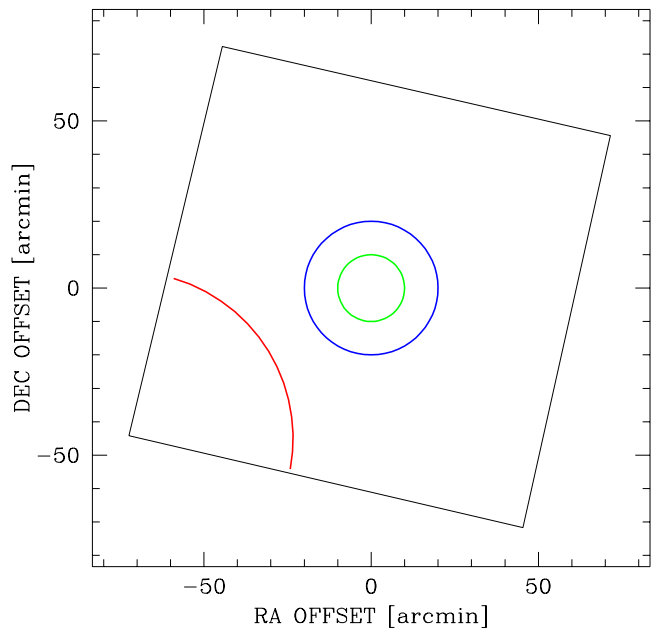,width=9cm} }
 
  \footnotesize
\caption[rms area]{Regions of different average rms noise defined for
  source extraction.  The radii of the circles are $10$, $20$ and $50$
  arcmin for the central inner, the central outer and corner circles
  respectively, and the corresponding rms noise is 40 $\mu$Jy, 60 $\mu$Jy
  and 100 $\mu$Jy. The rest of the field has a constant rms of 80 $\mu$Jy.}
\label{fig_rmsarea}
\end{figure} 
\begin{figure} 
\centerline{\psfig{figure=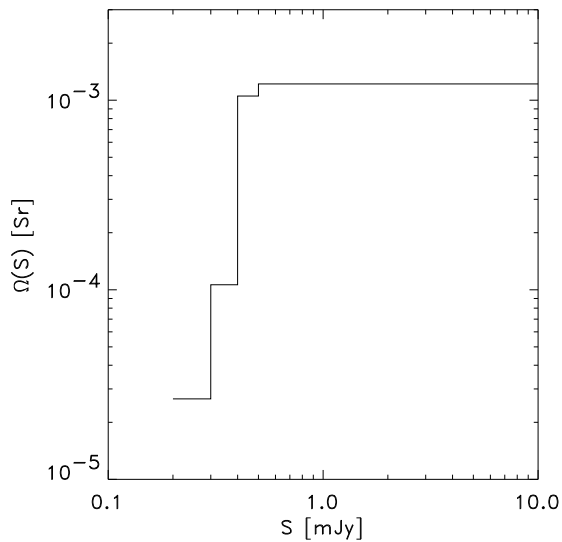,width=11cm} }
 
  \footnotesize
\caption[areal coverage]{Areal coverage of S1 field represented by the
  solid angle over which a source with peak flux $S$ can be detected.}
\label{fig_arealc}
\end{figure} 

\subsection{Noise distribution in the mosaic}

The amplitude distribution of pixel values in the final mosaic corresponds
to an almost Gaussian noise core plus a positive tail due to actual
sources. The standard deviation of the Gaussian which fits the distribution
is $\sim$80 $\mu$Jy, almost coincident with the theoretical rms value
expected in the image. However, as is clearly visible in
figure~\ref{fig_mosaic}, the bright, extended source PMNJ0042-4413 located
just outside the field of interest was strong enough (above 1 Jy) to
produce sidelobes and hence limit the dynamic range over a part of the
mosaic.

To deal with this problem, we analyzed the noise properties in the image as
a function of radial distance along two diagonals of the square defined by
the S1 field.  In figure~\ref{fig_noise} we plot the variations of the rms
noise with distance from the field center along the diagonal not affected
by the sidelobes of PMNJ0042-4413 (top panel) and along the one degraded by
them (bottom panel).  Except in the corner disturbed by PMNJ0042-4413,
where the rms noise is obviously higher than expected, the rms noise in the
mosaic is uniform over most of the image. It is smaller in the center where
a longer integration has been obtained, and almost flat over most of the
survey area.  Using figure~\ref{fig_noise} we defined regions with
different but uniform levels of rms noise to characterise the sensitivity
for source extraction. We adopted a 1$\sigma$ sensitivity of 40 $\mu$Jy in
a circular area of $10$ arcmin radius in the field center; 60 $\mu$Jy in a
$10 < r < 20$ arcmin annulus; 100 $\mu$Jy in the area delimited by a $50$
arcmin radius circle centered on the noisy corner and two sides of S1; and
80 $\mu$Jy over the rest of the field.  Figure~\ref{fig_rmsarea}
illustrates the location and extent of these regions.

Using these sensitivities we obtained the integral distribution of the rms
noise in the image and the detectability area as a function of flux density.
In figure~\ref{fig_arealc} the solid angle $\Omega(S)$ over which a source
with a peak flux density $S$ can be detected is plotted as a function of
flux density. The `step' structure of the areal coverage is due to the
fact that the field is divided in separate regions of constant rms noise.

\section{The source catalogue}

\subsection{Source detection}

\begin{figure}
\centerline{\psfig{figure=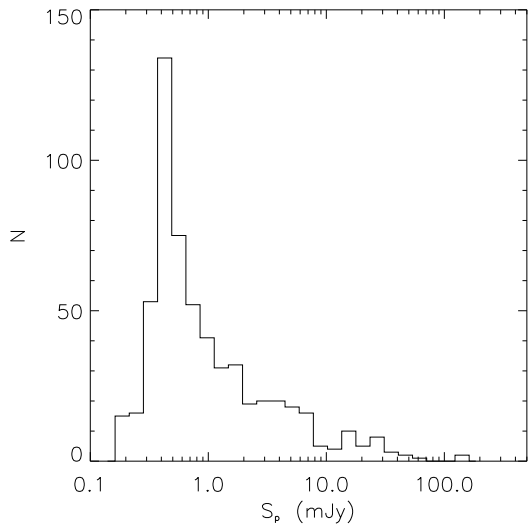,width=8cm}}

  \footnotesize
\caption[sp_histo]{Distribution of peak flux densities for radio sources
  of the S1 1.4 GHz complete sample.}
\label{sp_histo}
\end{figure}

The criterion for including a source in the catalogue is that its peak flux
density is $\geq 5$ times the average rms value in its region of the image.
Source extraction was performed using the {\sc miriad} task {\sc imsad}
(Image Search and Destroy), which searches for islands of pixels above a
given cutoff and attempts to fit Gaussian components to the islands. Source
parameters, derived by least-squares bi-dimensional Gaussian fitting, are
the right ascension and declination of the island centroid, peak flux
density, integrated flux density, deconvolved (from the beam) major axis
full width at half maximum FWHM (arcsec), deconvolved minor axis FWHM
(arcsec) and deconvolved position angle (degrees).

As discussed by Condon (1997), the results of Gaussian fitting can be
unreliable for sources with low signal-to-noise ratios. Thus, we used {\sc
  imsad} to extract all the sources whose peak flux, $S_p$, was greater
than 4 times the local rms value and then, for sources with $4\sigma < S_p
\leq 7\sigma$, we derived the peak flux density by second degree
interpolation (task {\sc maxfit}) and the total flux density by integrating
the image value in a rectangle around the source. Only sources with a {\sc
  maxfit} peak flux density $\geq 5\sigma$ were included in the final
sample.  For the other parameters (major axis, minor axis and position
angle) we retained the values given by {\sc imsad}.

\subsection{The catalogue}

\begin{figure*}
\vspace{-0.5cm}
\centerline{\psfig{figure=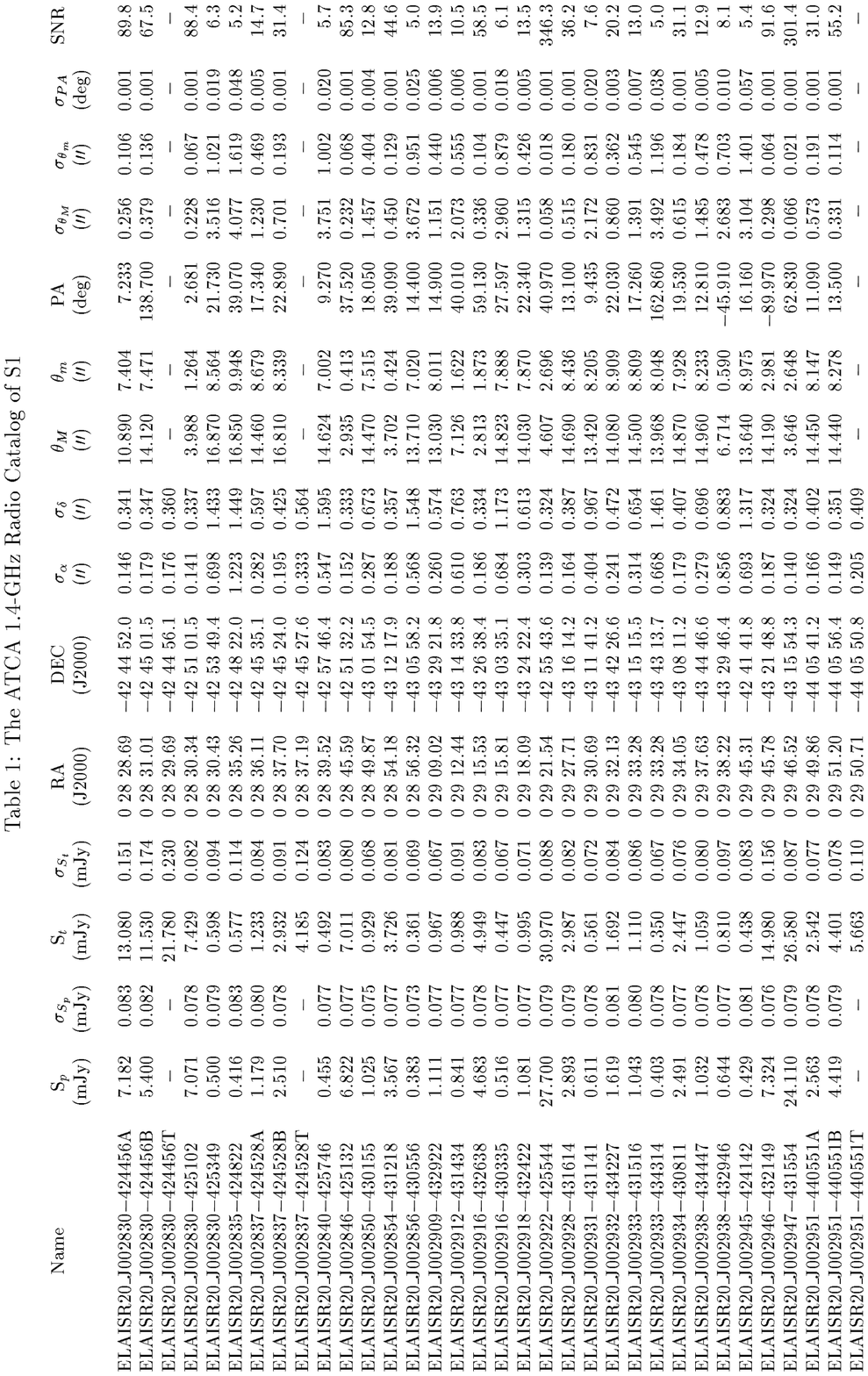,width=18cm}}
\addtocounter{table}{+1}
\end{figure*}

\begin{figure*}
\begin{minipage}{170mm}
  \centerline{ \psfig{figure=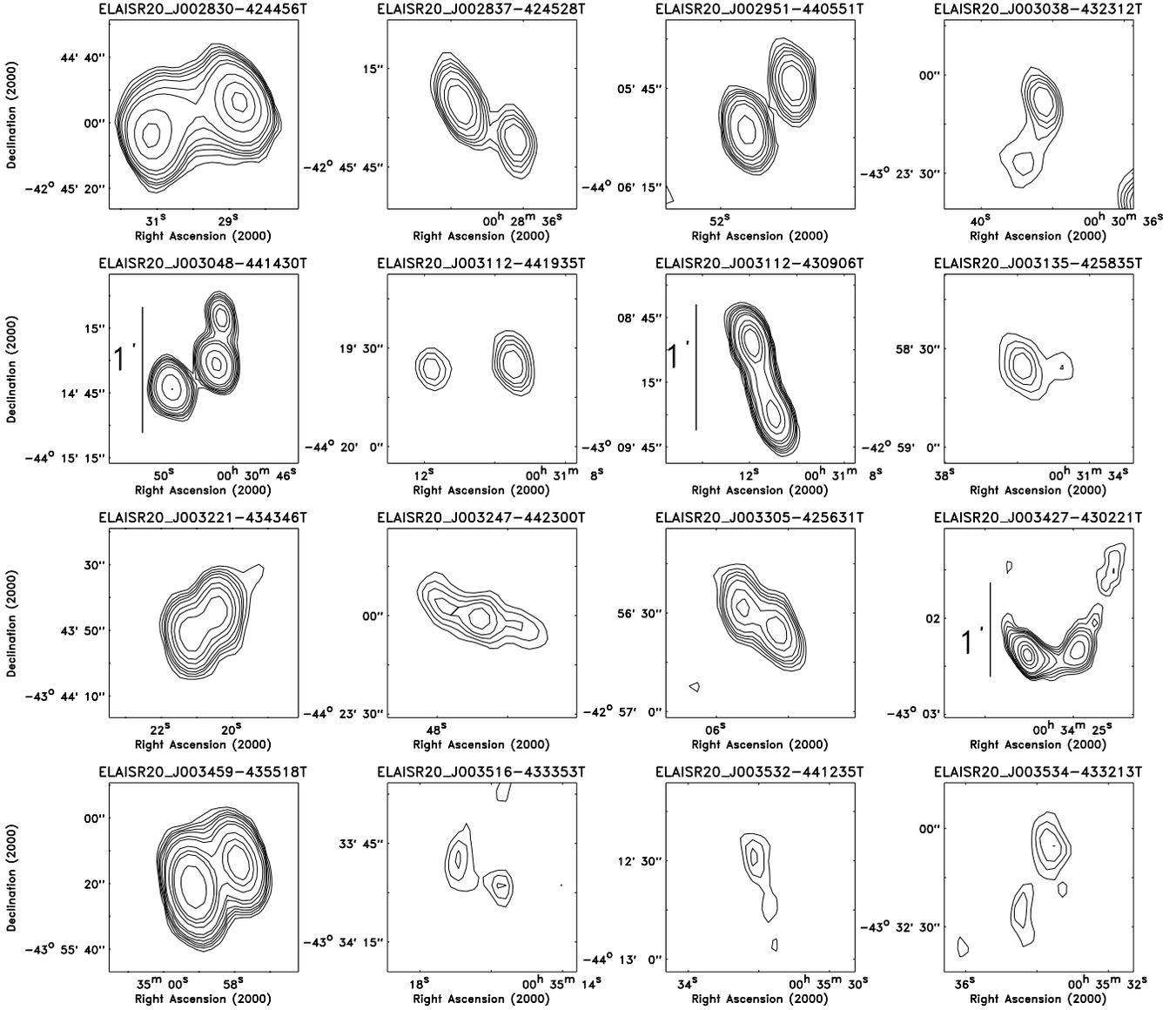,width=17cm} }
  \caption[contours]{Contour images of the 31 radio sources classified as
  double or multiple in S1. The contour levels are at 3, 4.5, 6, 7.5, 10,
  20, 40, 60, 80, 160, 320, 640 times the local rms value for all the
  sources.}
\label{fig_contour}
\end{minipage}
\end{figure*}

\begin{figure*}
\begin{minipage}{170mm}
  \centerline{ \psfig{figure=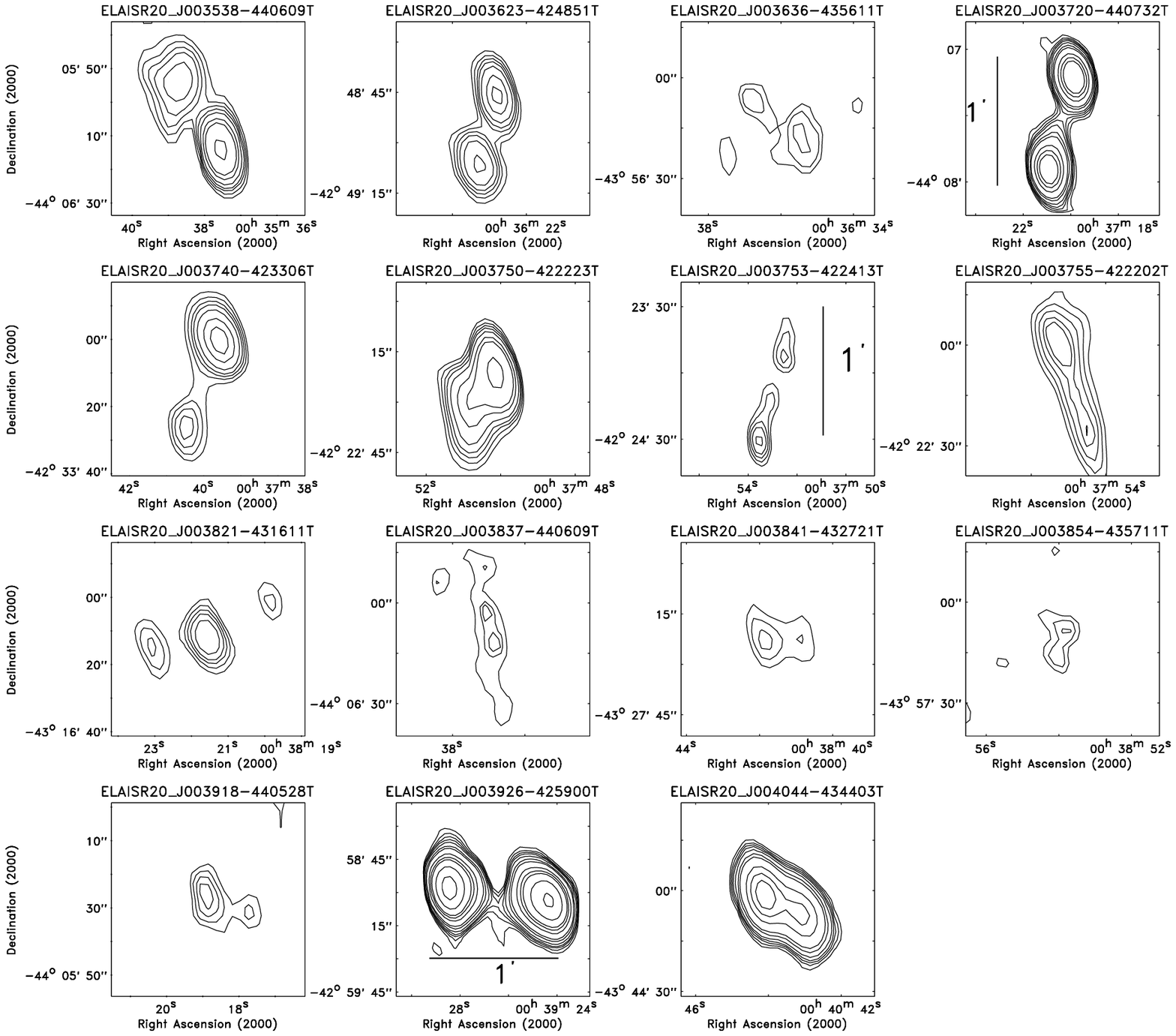,width=17cm} } 
  \contcaption{}
\end{minipage}
\end{figure*}

Our method selected 581 sources with peak flux density $\geq 5 \sigma$,
over an area of 4 deg$^2$.  Of these, 31 appear to have two or more
components.  Figure~\ref{sp_histo} illustrates the distribution of peak
flux density for the 581 sources in the catalogue. This distribution shows
that the majority of the radio sources in our sample ($\sim60$\%, 349/581)
have peak flux density in the sub-mJy region. Thus, our data-set provides a
large and statistically significant sub-sample of radio sources fainter
than 1 mJy.

We have considered as potential doubles all the sources separated by less
than about twice the FWHM of the synthesized beam ($\sim 30$ arcsec) and
having approximately equal flux densities in the two components (ratio
$\leq 2.5$).  For triple or multiple sources the distance between each
component and the probable nucleus has been considered, while the flux
density ratios have been computed between the various components excluding
the nucleus.  All the components satisfying the adopted criteria have been
assumed to form a unique, multiple source.  Figure~\ref{fig_contour} shows
contour images of the sources classified as double or multiple in the
catalogue.

The catalogue contains a total of 621 components, and reports the source
name, the peak flux density $S_p$ (in mJy), the integrated flux density
$S_t$ (in mJy), the source position (right ascension and declination at
equinox J2000), the FWHM of the major and minor axes, the positional angle
of the major axis, the signal-to-noise ratio and a character as a comment
about the deconvolution outcome (D=deconvolution of source size from beam
OK, P=deconvolution gave result close to point source, F=deconvolution
failed, E=multiple source). Whenever the deconvolution of the source size
from the beam failed (F), the fitted parameters (instead of the deconvolved
ones) are reported in the catalogue.  For double or multiple sources the
components are labeled `A', `B', etc., followed by a line labeled `T'
in which parameters for the total source are given. The position of the
total source has been computed as the flux-weighted average position for
all the components. In Table 1 the first page of the catalogue is shown as
an example.  The full catalogue will be available from
http://athena.ph.ic.ac.uk/.

\subsection{Errors in the source parameters}
The formal relative errors determined by a Gaussian fit are generally smaller than
the true uncertainties of the source parameters. Gaussian
random noise often dominates the errors in the data (Condon 1997). Thus, we used the
Condon (1997) error propagation equations to estimate the true errors on 
fluxes, axes and position angle:

\begin{equation}
\frac{\sigma^2_{S_p}}{S_p^2}=\frac{\sigma^2_{S_t}}{S_t^2}=\frac{\sigma^2_{\theta_M}}{\theta_M^2}=\frac{\sigma^2_{\theta_m}}{\theta_m^2}=\frac{\sigma_{PA}^2}{2} \left( \frac{\theta_M^2 - \theta_m^2}{\theta_M^2 \theta_m^2} \right)^2 = \frac{2}{\rho^2}
\end{equation}

\ni where $S_p$ and $S_t$ are the peak and the total fluxes, $\theta_M$ and $\theta_m$ the 
fitted FWHMs of the major and minor axes, PA is the position angle of the major axis
(the $\sigma$s are the relative errors) and $\rho$ is the signal--to--noise ratio, given by

\begin{equation}
\rho^2 = \frac{\theta_M \theta_m}{4\theta_N^2} \left[ 1+ \left( 
\frac{\theta_N}{\theta_M} \right)^2 \right]^{\alpha_M} \left[ 1+ \left(
\frac{\theta_N}{\theta_m} \right)^2 \right]^{\alpha_m} \frac{S_p^2}
{\sigma_{map}^2}
\end{equation}

\ni where $\sigma_{map}$ is the noise variance of the image and $\theta_N$ is
the FWHM of the Gaussian correlation length of the image noise ($\simeq$FWHM
of the synthesized beam). The exponents are $\alpha_M = 5/2$ and $\alpha_m
= 1/2$ for calculating $\sigma_M$, $\alpha_M = 1/2$ and $\alpha_m = 5/2$ for
calculating $\sigma_m$ and $\sigma_{PA}$ and $\alpha_M = \alpha_m = 3/2$ for
the flux densities.
\begin{figure} 
\centerline{\psfig{figure=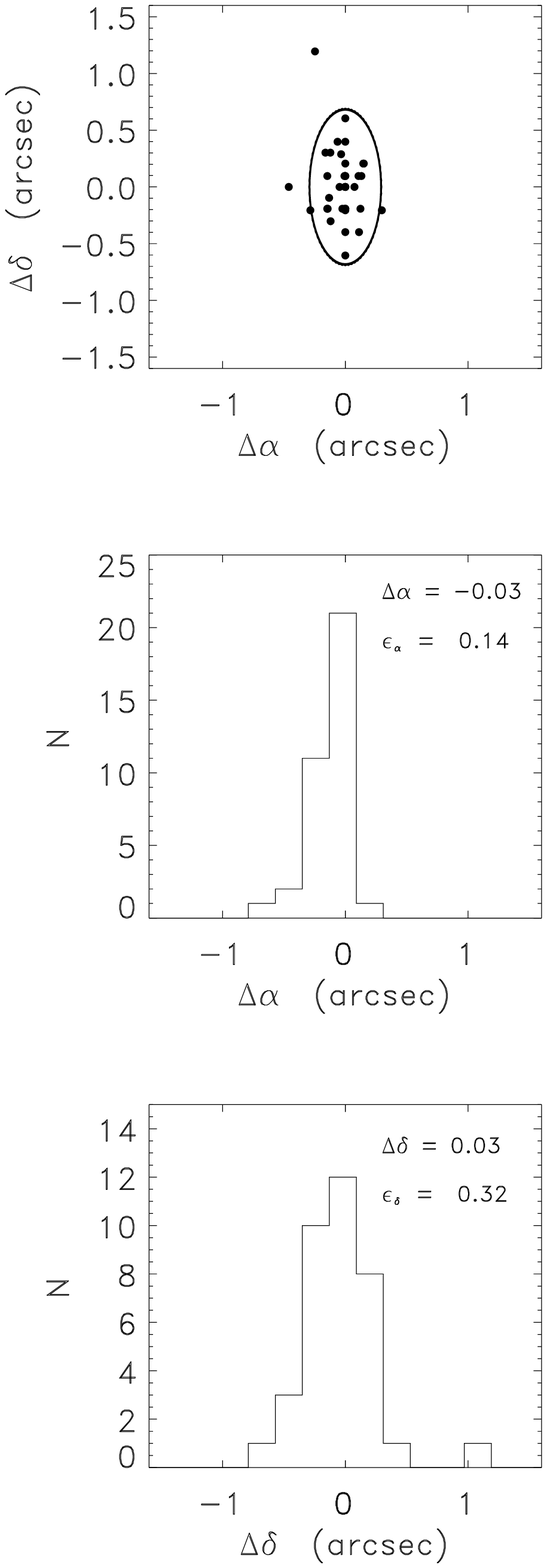,width=8.5cm} }
 
  \footnotesize
\caption[pos test bright]{Position errors for 36 bright single sources ($S > 10$ mJy) 
common to two individual pointing images composing the mosaic image. In the {\it top panel}
the semi-axes of the 90\% confidence ellipse shown are (2 ln 10)$^{1/2}$ times the rms
errors ($\varepsilon_{\alpha}$, $\varepsilon_{\delta}$) = ($0^{\prime \prime}.138$,
$0^{\prime \prime}.323$).}
\label{fig_posb}
\end{figure} 
\begin{figure} 
\centerline{\psfig{figure=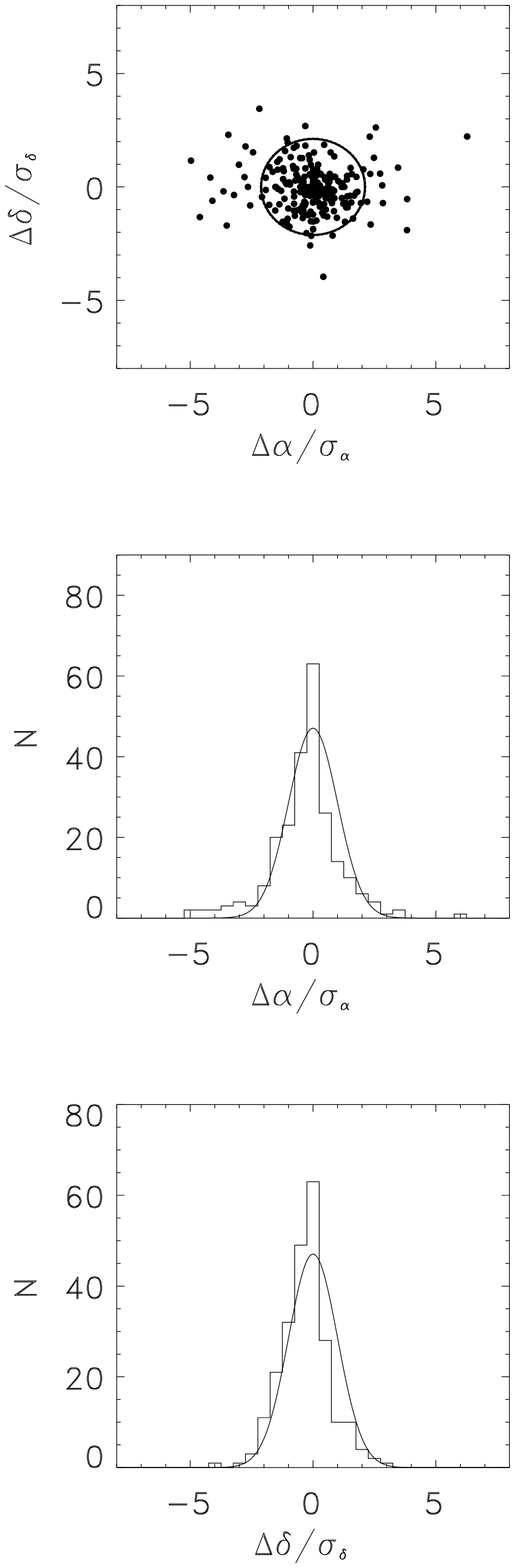,width=8.5cm} }
 
  \footnotesize
\caption[pos test faint]{Position errors for 236 weak single sources (fainter than 10 
mJy) in the overlapping regions, in units of the combined position uncertainties
($\sigma_{\alpha}$, $\sigma_{\delta}$). In the {\it top panel} the 90\% confidence
error circle is plotted. The smooth curves in the {\it central} and {\it lower panels}
represent the expected Gaussian of zero mean and unit variance.}

\label{fig_posf}
\end{figure} 
\begin{figure} 
\centerline{\psfig{figure=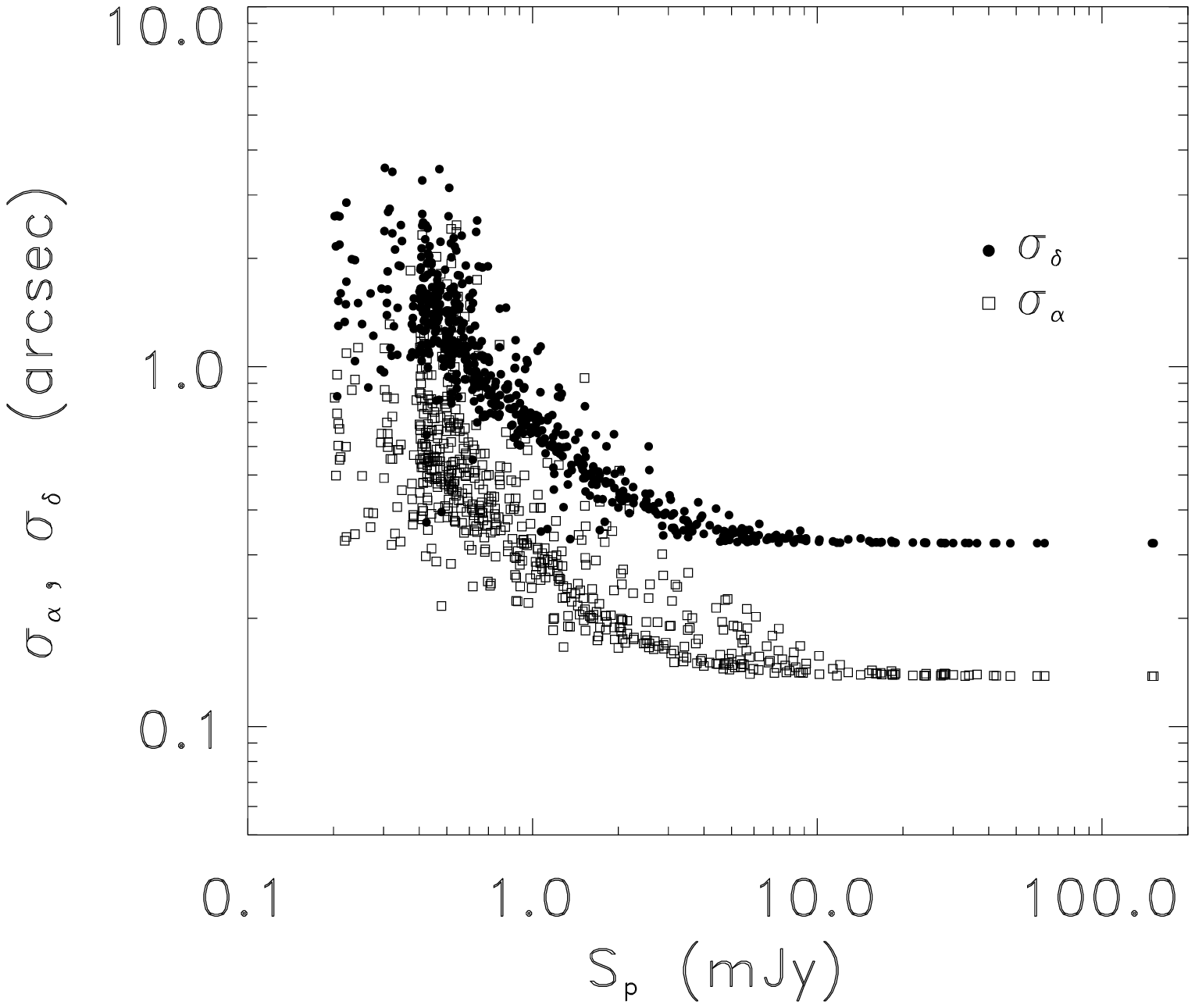,width=9cm} }
 
  \footnotesize
\caption[pos err]{Rms position uncertainties $\sigma_{\alpha}$ and $\sigma_{\delta}$
for all the single sources of peak flux $S_p$ in our catalogue.} 

\label{fig_poserr}
\end{figure}   
\ni These two equations are the master equations for estimating the variance in the 
parameters 
derived from a two--dimensional Gaussian fit.
The projection of the major and minor axis errors onto the right ascension and 
declination axes produces the total rms position errors given by Condon et al. (1998)

\begin{equation}
\sigma^2_{\alpha} = \varepsilon^2_{\alpha} + \sigma^2_{x_0} \sin^2(PA) + 
\sigma^2_{y_0} \cos^2(PA) \label{eq:ra_rms}
\end{equation}
\begin{equation}
\sigma^2_{\delta} = \varepsilon^2_{\delta} + \sigma^2_{x_0} \cos^2(PA) + 
\sigma^2_{y_0} \sin^2(PA) \label{eq:dec_rms}
\end{equation}

\ni where ($\varepsilon_{\alpha}, \varepsilon_{\delta}$) are the ``calibration'' 
errors, while $\sigma_{x_0}$ and $\sigma_{y_0}$ are $\theta_M^2/(4ln2)\rho^2$ 
and $\theta_m^2/(4ln2)\rho^2$ respectively.

Calibration errors cannot be determined from the survey image alone.
They can be determined from comparison with accurate positions of 
sources strong enough that the noise terms in equation~\ref{eq:ra_rms} 
and~\ref{eq:dec_rms} are much smaller than the calibration terms. 
Because S1 area is covered by no other 1.4-GHz radio catalogue 
suitable for estimating the calibration errors, we have used our data 
to derive the mosaic errors. This internal check gives only a lower limit
on the source position errors. However, Hopkins et al. (1998) have used 
Monte Carlo methods to explore source--fitting uncertainties in ATCA 
mosaiced images with statistical properties similar to those of the 
images discussed here. Their study implies that the positional errors are
rarely as large as 2 arcsec, and more typically lie below 1 arcsec.

The mosaic map contains several 
overlapping regions where sources are detected in two different pointings. 
We have run IMSAD on each single pointing map (after having corrected for 
the primary beam shape), considering the sources 
detected as an independent data set. Then we used the sources in common 
between two different pointing maps to estimate the positional errors. 
First we used 36 sources brighter than 10 mJy to calculate the mean image 
offsets and the mosaic uncertainties. 
Their offsets $\Delta\alpha$ and $\Delta\delta$ are shown in
figure \ref{fig_posb}. The mean offsets of our mosaic map are $<\Delta\alpha> =
-0.^{\prime \prime}027 \pm 0.^{\prime \prime}029$ 
and $<\Delta\delta> = 0.^{\prime \prime}031 \pm 0.^{\prime \prime}059$. Due to
the negligible values of these image offsets, the 
source positions in the catalogue have not been corrected for them. 
The offset distributions (reported in figure \ref{fig_posb})
indicate the mosaic position errors of our data, which are 
$\varepsilon_{\alpha} = 0.^{\prime \prime}138$ in right ascension and 
$\varepsilon_{\delta} = 0.^{\prime \prime}323$ in declination. 

Positional uncertainties for all sources have been calculated according
to equations~\ref{eq:ra_rms} and~\ref{eq:dec_rms}, adopting the mosaic errors 
derived using the
36 sources brighter than 10 mJy. To verify that these uncertainties are realistic,
especially at low flux densities where the noise and confusion components dominate the
errors, we used the 236 sources fainter than 10 mJy in the overlapping regions 
and found their position offsets. Figure~\ref{fig_posf} shows these offsets ($\Delta\alpha$,
$\Delta\delta$) normalized by the combined uncertainties ($\sigma_{\alpha}$, 
$\sigma_{\delta}$), where $\sigma_{\alpha}^2 = \sigma_{\alpha1}^2 + \sigma_{\alpha2}^2$
and $\sigma_{\delta}^2 = \sigma_{\delta1}^2 + \sigma_{\delta2}^2$ with $\sigma_{\alpha1}$,
$\sigma_{\alpha2}$, $\sigma_{\delta1}$ and $\sigma_{\delta2}$ being the errors on the single
measurements from equations~\ref{eq:ra_rms} 
and~\ref{eq:dec_rms}). As shown in figure~\ref{fig_posf}, where the expected Gaussians
are plotted over our normalized offset distributions, the normalized offset have nearly 
zero mean and unit rms scatter, 
verifying that our catalogue uncertainties are accurate also for weak sources.  
Many of these sources are slightly extended, so our catalogue positional errors for
weak sources include possible offsets between the source centroids and cores. 
In extended sources, there might be an intrinsic offset between the fitted 
centroid position and the real radio core position.

The rms position uncertainties ($\sigma_{\alpha}$, $\sigma_{\delta}$) of
all the sources in our catalogue are plotted as functions of peak flux in 
figure \ref{fig_poserr}.

\section{Source counts}

The sample of 581 sources with $S_p \geq 5\sigma$ has been used to
construct the source counts distribution. Complex sources or sources with
multiple components have been treated as a single radio source. Every
source was weighted for the reciprocal of its detectability area
(figure~\ref{fig_arealc}), defined as the area over which the source could
have been seen above the adopted threshold of $5\sigma$ (Katgert et al.
1973).

\begin{figure} 
\centerline{\psfig{figure=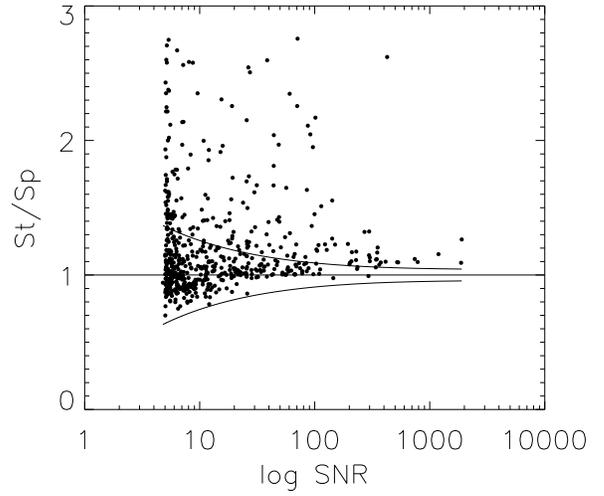,width=9cm} }
 
  \footnotesize
  \caption[st_sp]{The measured ratio of integrated to peak flux as a
    function of signal-to-noise ratio for the S1 ATCA survey. The upper
    line defines the upper envelope of the $S_t/S_p$ distribution
    containing the sources that we have considered unresolved.}

\label{fig_stsp}
\end{figure} 

\begin{table*}
\centering
\begin{minipage}{100mm}
  \caption{The 1.4 GHz ATCA Radio Source Counts}
  \label{counts_tab}

\begin{tabular}{cccccc}
& & & & \\ 
$S$ & $< S >$ & $N_s$ & $n=dN/dS$ & $nS^{2.5}$ & $N(>S$) \\ 
& & & & &  \\
 (mJy) & (mJy) & & (sr$^{-1}$ Jy$^{-1}$) & (sr$^{-1}$ Jy$^{1.5}$) & (sr$^{-1}$) \\ 
& & & & &  \\
  0.20--0.36 &  ~0.27 &  29 &4.53$\times10^9$& ~~5.3 $\pm$ ~~1.0&  7.24$\times10^5$ \\
  0.36--0.65 &  ~0.48 & 182 &1.50$\times10^9$& ~~7.7 $\pm$ ~~0.6&  4.31$\times10^5$ \\
  0.65--1.17 &  ~0.87 & 128 &2.56$\times10^8$& ~~5.7 $\pm$ ~~0.5&  1.32$\times10^5$ \\
  1.17--2.10 &  ~1.57 &  84 &7.40$\times10^7$& ~~7.2 $\pm$ ~~0.8&  6.91$\times10^4$ \\
  2.10--3.78 &  ~2.82 &  49 &2.39$\times10^7$& ~10.1 $\pm$ ~~1.4&  4.02$\times10^4$ \\
  3.78--6.80 &  ~5.07 &  42 &1.14$\times10^7$& ~20.9 $\pm$ ~~3.2&  3.45$\times10^4$ \\
  6.80--12.2 &  ~9.13 &  25 &3.77$\times10^6$& ~30.0 $\pm$ ~~6.0&  2.05$\times10^4$ \\
  12.2--22.0 &  ~16.5 &  18 &1.51$\times10^6$& ~52.2 $\pm$ ~12.3&  1.48$\times10^4$ \\
  22.0--39.7 &  ~29.6 &  15 &6.98$\times10^5$& 105.0 $\pm$ ~27.1&  1.23$\times10^4$ \\
  39.7--71.4 &  ~53.2 &   5 &1.29$\times10^5$& ~84.5 $\pm$ ~37.8&  4.10$\times10^3$ \\
~71.4--128.5 &  ~95.8 &   3 &4.31$\times10^4$& 122.4 $\pm$ ~70.7&  2.46$\times10^3$ \\
128.5--231.4 &  172.5 &   2 &1.60$\times10^4$& 197.1 $\pm$ 139.4&  1.64$\times10^3$ \\
& & & & \\ 
\end{tabular}
\end{minipage}
\end{table*}

An estimate of the extension of a source can be obtained from the ratio of
the integrated flux density to the peak flux density $S_t/S_p$.
Figure~\ref{fig_stsp} shows that the ratio has a skew distribution, with a
tail towards high flux density ratios, especially when the signal-to-noise
ratio is low.  To establish a criterion for extendedness, we have
determined the upper envelope of the distribution of $S_t/S_p$ containing
the unresolved sources.  The solid lines drawn in figure~\ref{fig_stsp}
represent the best-fit curves to the upper and lower envelopes of the band
containing all the sources considered to be unresolved. For these sources
we have adopted the peak flux in computing the source counts, while for all
the others, lying above the upper envelope, we have adopted the total flux.

In Table~\ref{counts_tab} the 1.4 GHz source counts are presented. The
columns give the adopted flux density intervals, the average flux density
in each interval computed as the geometric mean of the two flux limits, the
observed number of sources in each flux interval, the differential source
density (in sr$^{-1}$ Jy$^{-1}$), the normalized differential counts
$nS^{2.5}$ (in sr$^{-1}$ Jy$^{1.5}$) with estimated errors (as
$n^{1/2}S^{2.5}$) and the integral counts (in sr$^{-1}$).

\begin{figure} 
\centerline{\psfig{figure=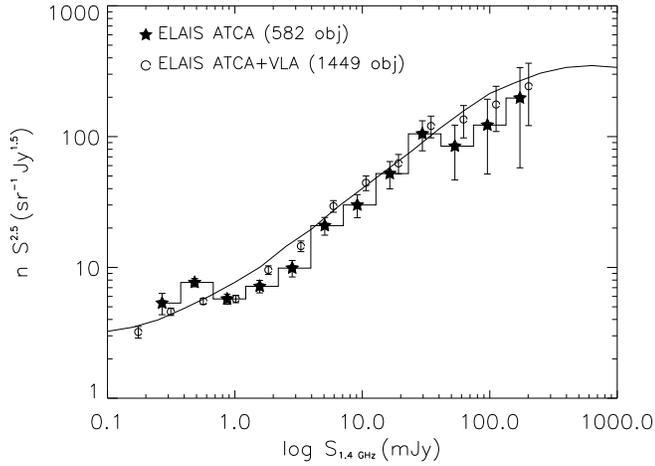,width=9cm} }
 
  \footnotesize
  \caption[counts]{The 1.4 GHz normalized differential source counts for 
    the ELAIS ATCA data. The abscissa gives the flux density (mJy) and the
    ordinate gives the differential number of sources normalized by
    $S^{2.5}$ (sr$^{-1}$ Jy$^{1.5}$). The solid line represents the model
    of Windhorst, Mathis \& Neuschaefer (1990) obtained by fitting counts
    from 24 different 1.4 GHz surveys. The filled stars are the counts
    obtained from our ATCA data in the ELAIS southern region, while the
    open circles are the total radio counts in the ELAIS regions obtained
    by combining the S1 data with the VLA data in the northern ELAIS
    regions (Ciliegi et al. 1998).}

\label{fig_counts}
\end{figure} 

The 1.4 GHz differential source counts of the ATCA data, normalized to
those expected in a Euclidean geometry by dividing by $S^{-2.5}$, are shown
in figure~\ref{fig_counts} (filled stars).  For comparison, source counts
from other surveys are also plotted. The solid line represents the model of
Windhorst, Mathis \& Neuschaefer (1990) obtained by fitting counts from 24
different 1.4 GHz surveys, while the open circles are the total counts
obtained by combining the S1 data with the VLA data in the northern ELAIS
regions (Ciliegi et al. 1998).  Our counts agree with those obtained by
previous surveys and confirm the upturn observed by several authors below
about 1mJy, considered as the characteristic feature of the `sub-mJy
population'.

A maximum likelihood fit to both our ATCA and ATCA+VLA 1.4 GHz counts with
two power laws:
\begin{equation} 
 \frac{dN}{dS} \propto \left\{ \begin{array}{ll}
                   S^{-\alpha_1} & \mbox{if $S>S_b$} \\
                   S^{-\alpha_2} & \mbox{if $S<S_b$}
                   \end{array}
             \right.  
\end{equation}
 
\ni gives the following parameters: $\alpha_1 = 1.73 \pm 0.11$, $\alpha_2 =
3.04 \pm 0.27$, $S_b =$($0.72 \pm 0.23$) mJy and $\alpha_1 = 1.74 \pm
0.06$, $\alpha_2 = 2.48 \pm 0.11$, $S_b =$($0.53 \pm 0.13$) mJy,
respectively. Although the errors for the ATCA data only are relatively
large, in both cases our best fit parameters suggest that the re-steepening
of the integral counts toward an Euclidean slope starts just below $\sim$1
mJy, in agreement with the recent results of Gruppioni et al.  (1997) and
Ciliegi et al. (1998). By contrast, in previous work (Windhorst et al.
1985, 1990), a best-fit value of 5 mJy had been derived for the flux
density at which the change in slope of the source counts occurs.

\section{Conclusions and future plans}
This paper presents a 1.4 GHz survey obtained with the ATCA of the European
Large Area ISO Survey (ELAIS) region S1 located in the southern celestial
hemisphere. A mosaic of forty-nine separate observations with different
pointing positions has provided an image with low and uniform rms noise
over the whole S1 area ($\sim$4 deg$^2$). The lowest $5\sigma$ flux density
reached by our observation is 0.2 mJy, while the bulk of the imaged area
has a 5$\sigma$ flux density limit of 0.4 mJy.

The observations provide a large, complete sample of 581 sources (with peak
flux greater than $5\sigma$), the majority of which have flux densities in
the sub-mJy range.  We have constructed the differential source counts over
the flux range 0.2--200 mJy. Our counts agree with the results from other
deep 1.4 GHz surveys (Windhorst et al.  1990; Gruppioni et al. 1997;
Ciliegi et al. 1998) and confirm the change in slope observed below $\sim$1
mJy.

This survey of the southern ELAIS field, S1, will be combined with the VLA
survey of the northern fields, N1, N2 and N3 (Ciliegi et al. 1998), and
deep multi-waveband data (optical and near-, mid- and far-infrared) in the same
area, to investigate the nature of the sub-mJy population. Moreover, due to
the accurate radio positions, the radio sample will play a crucial role in
the optical identification phase of the ISO program.

Spectroscopic observations for about 350 radio/ISO sources in S1 have recently
been obtained (end of September 1998) with the Anglo Australian Telescope 
(AAT) Two Degrees Field Spectrograph (2dF). These observations will 
provide crucial information about the spectroscopic nature and the 
redshift distribution of our objects, and allow estimation of the obscured 
star formation rate with two highly complementary samples, selected in the 
infrared and radio within the same volume of Universe.  In fact, since the 
radio luminosity traces the supernovae associated with star formation 
regions and is not affected by dust obscuration, independent estimates of 
the star formation rate for sources with reliable distance indications 
can also be calculated on the basis of radio flux density measurements 
(see Oliver et al.  1998).

\section*{Acknowledgments}
This paper is based on observations collected at the Australia Telescope
Compact Array (ATCA), which is funded by the Commonwealth of Australia for
operation as a National Facility by CSIRO. CG thanks Isabella Prandoni for
useful suggestions on mosaic data reduction strategies and Neil Killeen for
assistance in using the ATNF facilities at Epping and for kindly providing
the data tapes in a readable form. LC and AH acknowledge support from the
Australia Research Council. This work was supported by the EC TMR Network
program FMRX--CT96--0068.

\label{lastpage}
\end{document}